\documentclass[sn-nature]{sn-jnl}

\usepackage{graphicx}%
\usepackage{multirow}%
\usepackage{amsmath,amssymb,amsfonts}%
\usepackage{amsthm}%
\usepackage{mathrsfs}%
\usepackage[title]{appendix}%
\usepackage{xcolor}%
\usepackage{textcomp}%
\usepackage{manyfoot}%
\usepackage{booktabs}%
\usepackage{algorithm}%
\usepackage{algorithmicx}%
\usepackage{algpseudocode}%
\usepackage{listings}%
\usepackage{subcaption}
\usepackage{color,soul}
\usepackage[T1]{fontenc}
\usepackage{changepage}
\def\jcm{J/cm$^2$}

\begin{document}

\title[Grayscale control of local magnetic properties with direct-write laser annealing]{Grayscale control of local magnetic properties with direct-write laser annealing}

\author*[1,2]{\fnm{Lauren J.} \sur{Riddiford}}\email{lauren.riddiford@psi.ch}
\equalcont{These authors contributed equally to this work.}

\author*[1,2]{\fnm{Jeffrey A.} \sur{Brock}}\email{jeffrey.brock@psi.ch}
\equalcont{These authors contributed equally to this work.}

\author[1,2]{\fnm{Katarzyna} \sur{Murawska}}

\author*[1,2]{\fnm{Ale\v{s}} \sur{Hrabec}}\email{ales.hrabec@psi.ch}

\author[1,2]{\fnm{Laura J.} \sur{Heyderman}}

\affil[1]{\orgdiv{Laboratory for Mesoscopic Systems}, \orgname{Department of Materials, ETH Zurich}, \orgaddress{\postcode{8093}, \city{Zurich}, \country{Switzerland}}}

\affil[2]{\orgdiv{Laboratory for Multiscale Materials Experiments}, \orgname{Paul Scherrer Institute}, \orgaddress{\postcode{5232}, \city{Villigen PSI}, \country{Switzerland}}}

\abstract{Across the fields of magnetism, microelectronics, optics, and others, engineered local variations in physical properties can yield groundbreaking functionalities that play a crucial role in enabling future technologies \cite{choe2019ion, ghosh2020demand, bonfadini2022femtosecond, sarcan2023_FIBoptics,chitnis2011_microfluidics, kiechle2023spin}. Beyond binary modifications, 1D lateral gradients in material properties (achieved by gradients in thickness, stoichiometry, temperature, or strain) give rise to a plethora of new effects in thin film magnetic systems \cite{yamahara2021flexoelectric,kim2023chiral,hou2016observation,uchida2008observation,yu2014switching}. However, extending such gradient-induced behaviors to 2D is challenging to realize with existing methods, which are plagued by slow processing speeds, dose instabilities, or limitation to variation along one dimension  \cite{gigax2015_ionspeed,schmidt_variableion,mcginn2019_combinatorial}. Here, we show for the first time how commonplace direct-write laser exposure techniques, initially developed for grayscale patterning of photoresist surfaces \cite{grushina2019_DWLgrayscale}, can be repurposed to perform grayscale direct-write laser annealing. With this technique, we demonstrate the ease with which two-dimensional, continuous variations in magnetic properties can be created at the mesoscopic scale in numerous application-relevant materials, including ferromagnetic, ferrimagnetic, and synthetic antiferromagnetic thin-film systems. The speed, versatility, and new possibilities to create complex magnetic energy landscapes offered by direct-write laser annealing opens the door to the lateral modification of the magnetic, electronic, and structural properties of a variety of thin films with an abundance of applications.}

\maketitle

Laser annealing, in which thermal energy is imparted to a material using a high-powered laser, has been used in magnetism to modify the saturation magnetization, exchange bias, and magnetic easy axis orientation of thin films \cite{ando1983localised, berthold2014exchange, sharma_exchange_2019}. However, the instruments previously used have been custom setups or repurposed lasers with a spot size of $\sim10-20$ $\mu$m and without spatial patterning capabilities beyond that provided by a simple Gaussian intensity profile \cite{levati2023phase}. While local control of magnetic anisotropy through electron beam exposure has been demonstrated, the technique has been applied sparingly \cite{allenspach_local_1998, guang_electron_2020}.

Here, we demonstrate the implementation of direct-write laser annealing (DWLA) to create arbitrarily-shaped magnetic energy landscapes in numerous important magnetic thin-film systems. We locally engineer continuous variations in the magnetic anisotropy in ferromagnetic and synthetic antiferromagnetic materials, the compensation temperature of ferrimagnets, and the Ruderman-Kittel-Kasuya-Yosida (RKKY) coupling strength of synthetic antiferromagnets, exemplifying the broad utility of DWLA. Crucially, as DWLA does not require patterned resist layers or ultrahigh vacuum environments, the process is significantly streamlined compared to other approaches. In addition, the applicability of this technique to a wide variety of material systems opens the door to creating novel and precise potential energy landscapes for spintronic and magnonic applications, and beyond. 

The principle of DWLA is summarized in Fig. \ref{fig:CFBvPower}a. A 3D CAD structure (input) is used to create gradients in the film plane, with the local laser fluence determined by the height of the 3D structure. The photolithography software interprets this 3D object as a grayscale image, where white indicates maximum laser power and black indicates no exposure. The laser writer is then used to expose the pattern onto the film by rastering the laser across the surface and modulating the laser power according to the design. The resulting magnetic texture is observed with a Kerr microscope (output), seen in Fig. \ref{fig:CFBvPower}b. In the following sections, we highlight four material systems that illustrate four mechanisms of thermally-induced material transformation that DWLA offers: crystallization, oxidation, interfacial alloying, and diffusion in multilayers. 

\section{Interface crystallization in ferromagnetic heterostructures}
CoFeB/MgO-based heterostructures have been extensively investigated for implementation in magnetic tunnel junctions (MTJs) due to the high tunneling magnetoresistance of CoFeB/MgO/CoFeB. Furthermore, strong perpendicular magnetic anisotropy (PMA) can be achieved in these heterostructures through thermal annealing, where crystallization of the CoFeB/MgO layers yields interfacial PMA ($K_\text{i}$) due to hybridization of the Fe-O orbitals. The effective magnetic anisotropy energy $K_{\text{eff}}$ is given by: 
\begin{equation}
    K_{\text{eff}} = K_\text{u} - \frac{1}{2} \mu_0 M_\text{s}^2 + \frac{K_\text{i}}{t_\text{CFB}}
    \label{eq:fm_energy}
\end{equation}
where $K_\text{u}$ is the (negligible) magnetocrystalline anisotropy,  $\mu_0$ is the vacuum permeability, $\mu_0 M_\text{s}$ is the saturation magnetization, $\frac{1}{2} \mu_0 M_\text{s}^2$ is the shape anisotropy, and $t_\text{CFB}$ is the CoFeB thickness \cite{liu_effect_2016}. By tuning the magnitude of $K_\text{i}$, the effective anisotropy can be modified from in-plane ($K_{\text{eff}}<0$) to out-of-plane ($K_{\text{eff}}>0$). In particular, when $t_\text{CFB}\leq1.4$ nm, interfacial PMA can overcome the shape anisotropy. With DWLA, thermal energy is deposited in small regions of CoFeB with high grayscale precision, allowing for continuous tuning of $K_{\text{eff}}$ over a broad range. In this materials system, we observe that DWLA increases the ordering of the film, providing a process window where the properties are transformed through crystallization rather than film damage.

\begin{figure}[htb!]
    \centering
    \includegraphics[width=\textwidth]{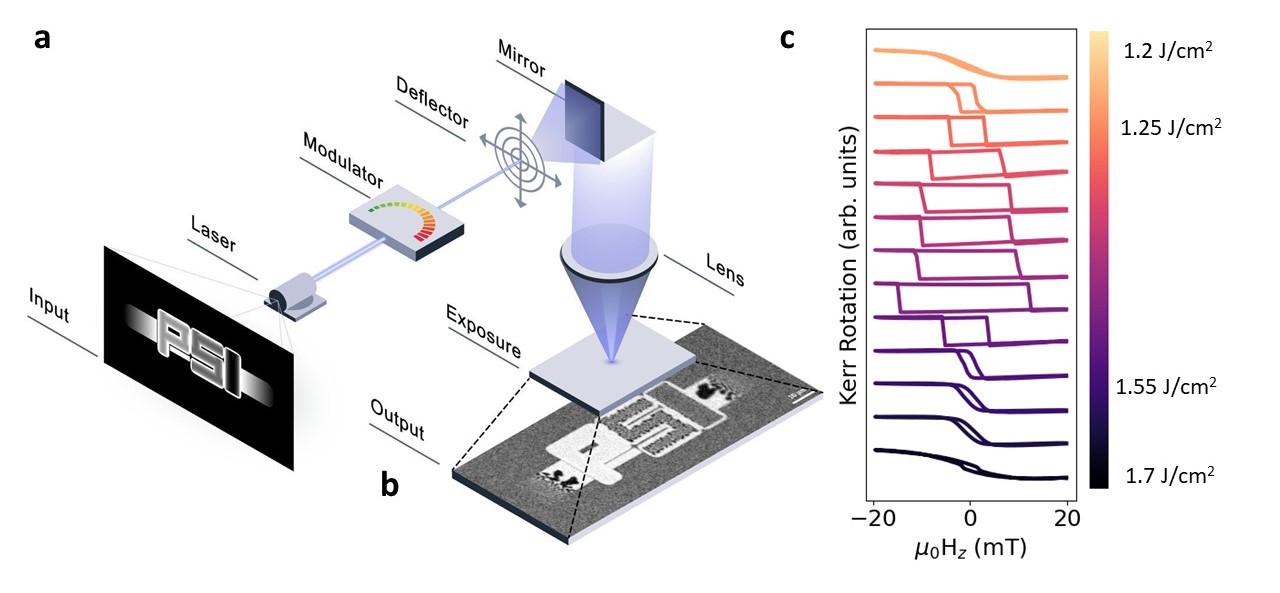}
    \caption{\textbf{DWLA and local anisotropy tuning of ferromagnetic thin films.} \textbf{a,} Procedure for local grayscale laser annealing. 2D grayscale intensity in the pattern is interpreted as a laser fluence adjustment, which is executed by the modulator. The deflector rasters the laser on the film as the stage simultaneously moves to write a 2D pattern. \textbf{b,} A background-subtracted Kerr micrograph of the written pattern on a CoFeB thin film (scale bar = 10 $\mu$m). The magnetic domain size increases as laser fluence increases, indicating stronger PMA. \textbf{c,} Changes in the pMOKE hysteresis loops for regions of 1.3-nm-thick CoFeB exposed to increasing laser fluence. PMA is achieved in a fluence range of 1.25-1.55 \jcm, while fluences >1.55 \jcm\ disrupt PMA. The standard error of the measurements shown in c are smaller than the lines used.}\label{fig:CFBvPower}
\end{figure}

 Ta/CoFeB/MgO films are grown with in-plane magnetic anisotropy. As regions of the film are laser annealed with increasing fluence, hysteresis loops recorded with polar magneto optic Kerr effect (pMOKE) magnetometry become increasingly square. As shown in Fig. \ref{fig:CFBvPower}c, sharp switching steps indicative of PMA are observed by a fluence of 1.25 \jcm. When the laser fluence is further increased above a threshold value of 1.55 \jcm, the hysteresis loops become more slanted, indicating a loss of PMA. These results are consistent with those obtained by furnace annealing CoFeB films, where there is an optimal temperature range of $\sim200-350$ $^{\circ}$C that promotes strong PMA \cite{liu_effect_2016}. When furnace annealing above $\sim350$ $^{\circ}$C, Ta diffusion degrades the interfacial anisotropy energy, and the magnetization easy axis falls in-plane. Comparing the effects of DWLA to furnace annealing, we estimate that DWLA in the optimal fluence range is equivalent to 1 hour in a 300 $^{\circ}$C furnace (see Extended Data Fig. 1). Meanwhile, DWLA accomplishes the same effect for a $100\times100$ $\mu$m$^2$ square in 30 seconds. An anisotropy energy range of up to $\Delta K_\text{eff}\approx3\times10^5$ J/m$^3$ can be created in arbitrary patterns in the film plane (Fig. \ref{fig:CFBvPower}b). This energy difference is sufficient to dramatically influence propagation of spin waves \cite{yu2023magnon}. Furthermore, DWLA could be used to selectively anneal a series of MTJs to different PMA strengths for the construction of sophisticated neural networks \cite{rzeszut2022multi}.

\section{Control of ferrimagnetic compensation temperature by oxidation}
Near the magnetic compensation temperature (T$_\text{m}$), it is well-known that transition metal-rare earth (TM-RE) ferrimagnetic films exhibit novel, application-relevant magnetization dynamics, such as extraordinarily efficient current-induced domain wall motion \cite{caretta_ferrimagnet,kim_ferrimagnet,blasing_ferrimagnet} and ultrafast laser-induced magnetization switching \cite{stanciu_aos}. More recently, it has been demonstrated that the handedness of spin waves in TM-RE ferrimagnets depends on whether the sample temperature is above or below T$_\text{m}$ \cite{kim2020distinct}. In addition to the demonstration of localized, binary control of T$_\text{m}$ using focused ion irradiation and selective oxidation techniques \cite{liu2023strong,lee_ferrimagnet_irradiation,frkackowiak_ferrimagnet_irradiation}, the symmetry breaking originating from 1D lateral gradients in T$_\text{m}$ has recently been shown to be a valuable means of achieving magnetic field-free, all-electrical magnetization switching \cite{lee_ferrimagnet_irradiation}. As all-electrical control of magnetization is necessary to realize energy-efficient, high-density spintronic computing architectures, there is a crucial need to develop scalable techniques by which these functionalities can be extended to 2D \cite{lin2019_2D}.

\begin{figure}[htb!]
    \centering
    \includegraphics[width=1\linewidth]{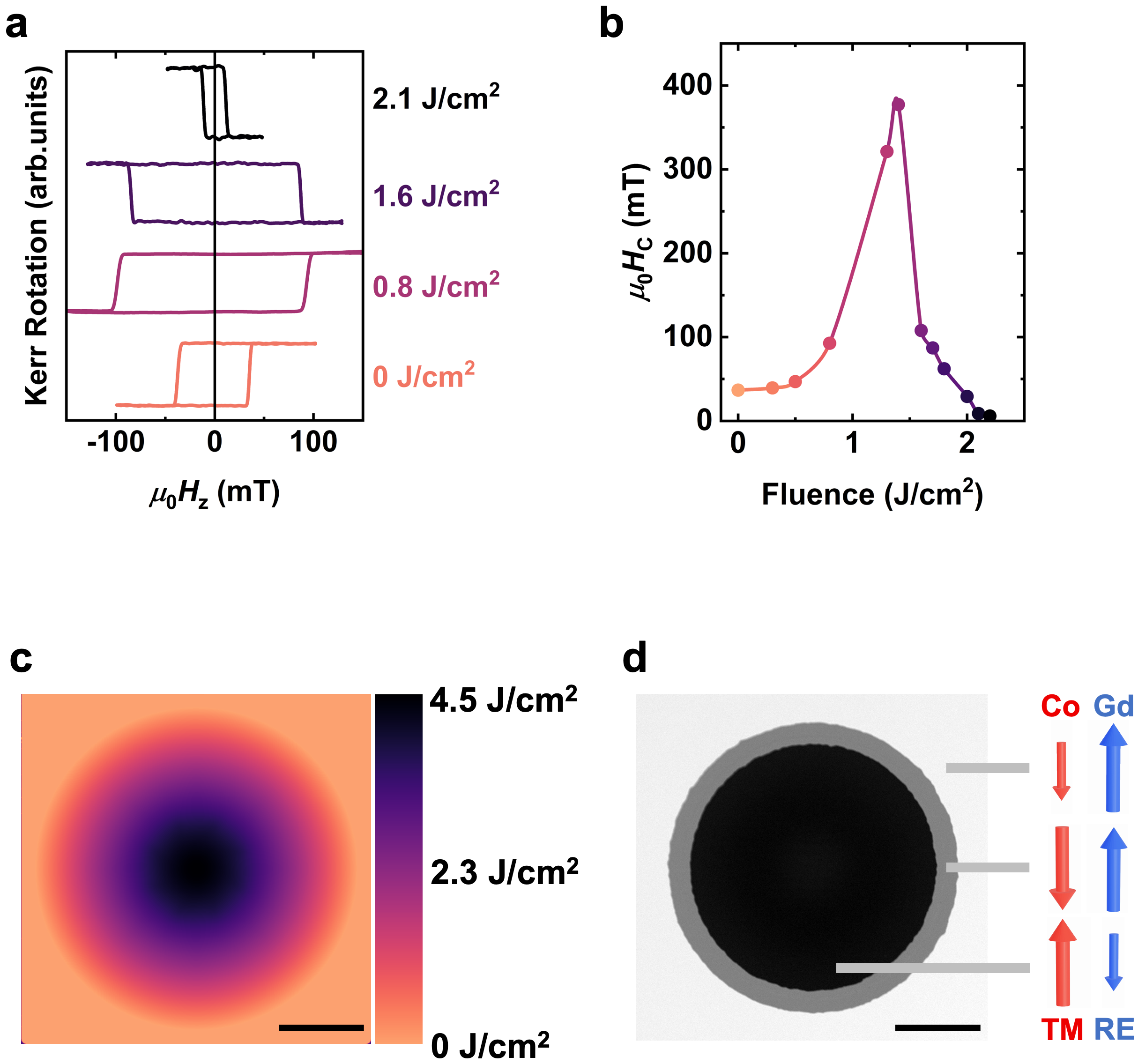}
   \caption{\textbf{Tuning the ferrimagnetic compensation temperature in CoGd films with DWLA.} \textbf{a,} pMOKE hysteresis loops for $100 \times100$ $\mu$m$^2$ regions that have been subjected to DWLA using different laser fluences. \textbf{b,} Coercive field $\mu_{0}H_\text{c}$ as a function of laser fluence extracted from pMOKE loops. \textbf{c,}  The 2D fluence profile used to create the ferrimagnetic ``compensation surface'' demonstrated in \textbf{d} (scale bar = 10 $\mu$m). \textbf{d,} pMOKE image of the compensation surface created using the fluence profile shown in \textbf{c} and schematic representations of the magnetic Co (red) and Gd (blue) sublattice configurations of the RE-dominant (top), compensated (middle), and TM-dominant (bottom) regions that define the compensation surface (scale bar = 10 $\mu$m). The measurement protocol used to generate \textbf{d} is provided in the Methods section. See Extended Data Fig. 2a for the detailed pMOKE characterizations used to generate \textbf{b}. The standard error of the measurements shown in \textbf{a} and \textbf{b} are smaller than the symbols or lines used.}
    \label{fig:Ferri_1}
\end{figure}

Here, we demonstrate that DWLA can be applied to TM-RE ferrimagnets to impart two-dimensional, continuous gradients in T$_\text{m}$ through local oxidation. For this, we exposed $100 \times100$ $\mu$m$^2$ regions of a CoGd film (see Methods for the sample composition) using varying laser fluences, and collected hysteresis loops using pMOKE (Fig. \ref{fig:Ferri_1}a). The positive polarity of the room temperature pMOKE hysteresis loop of a virgin CoGd film (orange loop in Fig. \ref{fig:Ferri_1}a) indicates the as-deposited film is RE dominant in its magnetic properties at room temperature \cite{khorsand_ferrimagnet_specific}. For laser fluences less than 1.3 \jcm, the coercivity increases with increasing fluence (light purple loop in Fig. \ref{fig:Ferri_1}a). For fluences greater than or equal to 1.3 \jcm, the polarity of the pMOKE loop is reversed relative to the virgin film (dark purple loop in Fig. \ref{fig:Ferri_1}a), and the coercivity decays with increasing fluence (black loop in Fig. \ref{fig:Ferri_1}a). The change in the pMOKE loop polarity, coupled with the non-monotonic, divergent trend in coercivity as a function of fluence shown in Fig. \ref{fig:Ferri_1}b demonstrates how DWLA permits tuning of magnetic compensation in ferrimagnetic films \cite{liu2023strong,caretta_ferrimagnet}. This scenario is further confirmed by magnetometry measurements of continuous films uniformly exposed using various laser fluences (see Extended Data Figs. 2b and 2c).

Unlike past approaches, DWLA allows for continuous 2D patterning such as the radial fluence gradient demonstrated in Fig. \ref{fig:Ferri_1}c. In Fig. \ref{fig:Ferri_1}d, we show how the fluence profile of Fig. \ref{fig:Ferri_1}c results in the formation of a two-dimensional “compensation surface” at remanence after the sample was exposed to a +500 mT perpendicular magnetic field, characterized by the presence of three distinct magnetic regions whose configurations are illustrated in the insets of Fig. \ref{fig:Ferri_1}d. Separating the TM (in black) and RE (in white) dominant regions, there exists a shell of intermediate gray contrast, corresponding to an area whose magnetization has not switched relative to the subtracted reference image (see Methods). Given that the coercivity diverges near magnetic compensation, this inability to switch the magnetic orientation in a large magnetic field indicates that these areas are close to magnetic compensation at room temperature. While linear composition gradients have previously been used to obtain 1D compensation walls \cite{hrabec_compwall}, DWLA permits this first-ever demonstration of a 2D ``compensation surface.''

\section{Modifying RKKY coupling fields in synthetic antiferromagnets through interfacial alloying}
Thin-film systems in which individual ferromagnetic layers are antiferromagnetically coupled to each other through a spacer layer that fosters indirect RKKY coupling (so-called synthetic antiferromagnets, or SAFs) have long enjoyed significant interest. Besides playing a crucial role in modern magnetic sensing technologies that operate on the principle of giant magnetoresistance \cite{grunberg_RKKY,baibich_RKKY,parkin_RKKY}, there has been a renewed enthusiasm for SAFs in the past decade due to their extraordinarily fast and efficient domain wall motion in response to a spin-orbit torque \cite{yang_SAFDW}. More recently, the utility of creating local gradients in the RKKY coupling field in SAFs has been highlighted by works showing that such gradients permit field-free, all-electrical magnetization switching using spin-orbit torques \cite{wang2023_safgradswitch}. As such, the development of new approaches by which the RKKY coupling strength of SAFs can be locally controlled is of paramount technological interest in the development of magnetic components for integration in spin-based computing and memory schemes \cite{lin2019_2D,mitin2018_AFMs,koch2020_SAFmanipulation}.

\begin{figure}[htb!]
    \centering
    \includegraphics[width=1\linewidth]{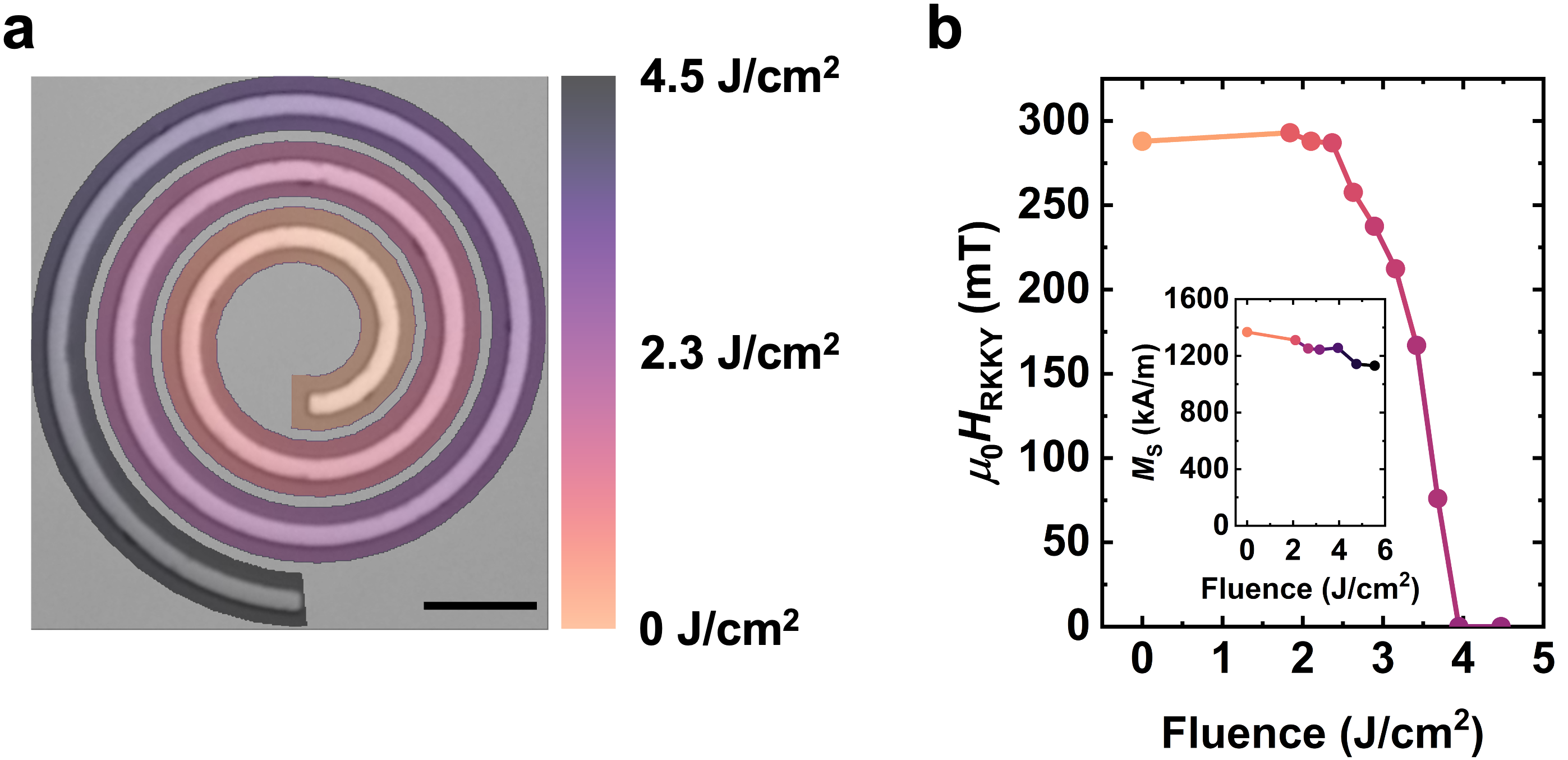}
    \caption{\textbf{Continuous reduction of the RKKY coupling field in Co/Cr/Co SAF structures with DWLA.} \textbf{a,} Optical micrograph of the spiral track structure with an overlay of the 2D fluence profile imparted to the structure (scale bar = 10 $\mu$m). \textbf{b,} RKKY coupling field ($\mu_{0}H_\text{RKKY}$) as a function of laser fluence, measured along the spiral structure shown in \textbf{a}. The inset shows the saturation magnetization $M_\text{s}$ as a function of laser fluence, determined from SQUID - VSM measurements. See Extended Data Fig. 3 for the full pMOKE and SQUID - VSM measurements used to generate Figs. 3\textbf{b}. The standard error of the measurements shown in \textbf{b} are smaller than the symbols or lines used.}
    \label{fig:SAFJB_2}
\end{figure}

To demonstrate the unique capability of DWLA to pattern the local RKKY coupling in SAFs by means of interfacial alloying, we consider a Co/Cr/Co multilayer (see Methods for the full sample structure). The as-deposited Co/Cr/Co sample exhibits a pMOKE hysteresis loop typical of SAFs with PMA (see Extended Data Fig. 3a). Defining the RKKY coupling field $\mu_{0}H_\text{RKKY}$ as the field about which the spin-flip transitions are centered, the as-deposited sample possesses a $\mu_{0}H_\text{RKKY}$ of 360 mT. Next, we patterned a spiral-shaped exposure gradient, as shown in Fig. \ref{fig:SAFJB_2}a. Starting from zero fluence, the fluence was continuously increased along the spiral to a maximum of 4.5 \jcm. From pMOKE hysteresis loops collected at selected locations along the spiral, we have extracted $\mu_{0}H_\text{RKKY}$ as a function of laser fluence, as shown in Fig. \ref{fig:SAFJB_2}b (see Extended Data Fig. 3a for the full pMOKE characterization). From Fig. \ref{fig:SAFJB_2}b, it can be seen that an increase in laser fluence causes a decrease in $\mu_{0}H_\text{RKKY}$. When Co/Cr/Co is exposed using laser fluences greater than approximately 4 \jcm, a spin-flip transition is no longer apparent in the pMOKE hysteresis loops (see Extended Data Fig. 3a), suggesting that the two FM layers comprising the SAF structure are no longer antiferromagnetically coupled to one another, to which we assign a $\mu_{0}H_\text{RKKY}$ value of zero.

To understand the relationship between laser fluence and the measured $\mu_{0}H_\text{RKKY}$, we consider the simple algebraic relationship: 
\begin{equation}
    J_\text{RKKY} =  t_\text{m} M_\text{s} \mu_0 H_\text{RKKY}
    \label{eq:rkky}
\end{equation}
t$_\text{m}$ is the total thickness of magnetic material in the SAF, and J$_\text{RKKY}$ is the interlayer coupling strength (set by the thickness of the spacer layer). Therefore, possible explanations for the observed reduction in $\mu_0 H_\text{RKKY}$ with increasing laser fluence could be a reduction in $J_\text{RKKY}$ (due to a change in the effective thickness or stoichiometry of the RKKY spacer layer) or $M_\text{s}$ (resulting from the oxidation of Co). To delineate between these two mechanisms, we determined the $M_\text{s}$ of continuous Co/Cr/Co films uniformly exposed using various laser fluences (see the inset of Fig. \ref{fig:SAFJB_2}b and Extended Data Fig. 3b). Over the range of fluences that modify the magnetic properties of our Co/Cr/Co SAFs (0 - 5.5 \jcm), DWLA reduces $M_\text{s}$ by approximately 15\% relative to the as-deposited sample. However, within the context of Eq. \ref{eq:rkky}, a 15\% reduction in $M_\text{s}$ cannot be solely responsible for the extensive changes in RKKY coupling field evident in Fig. \ref{fig:SAFJB_2}b. Thus, while a reduction in $M_\text{s}$ due to heating-induced oxidation of Co may occur, a change in the effective thickness of the Cr spacer layer as a result of the alloying of Cr and Co must also be responsible for the observed variation in both the strength and nature (\textit{i.e.}, ferromagnetic or antiferromagnetic) of the interlayer RKKY coupling.

\section{Diffusion-induced anisotropy modification of synthetic antiferromagnets}
In addition to local control of RKKY coupling, choice of different materials can allow us to instead modify the magnetic anisotropy of a SAF. Two CoFeB/Pt magnetic multilayers coupled by a Ru spacer are synthesized for this purpose. In previous studies of CoFeB/Pt multilayers, it was found that PMA is lost when the structure was annealed above $\sim300$ $^{\circ}$C \cite{zhu_thermal_2012}. With DWLA, thermally-induced diffusion within each magnetic layer leads to a coherent change in magnetic anisotropy of a SAF as the laser fluence increases -- from strongly out-of-plane, to weakly out-of-plane, to in-plane, without disrupting the RKKY coupling. This ability to modify anisotropy is of particular technological relevance for domain wall devices, as domain walls are expected to react more strongly to an anisotropy gradient in an antiferromagnet, and therefore in SAFs, than a ferromagnet \cite{wen2020ultralow}.

\begin{figure}[htb!]
    \centering
    \includegraphics[width=\textwidth]{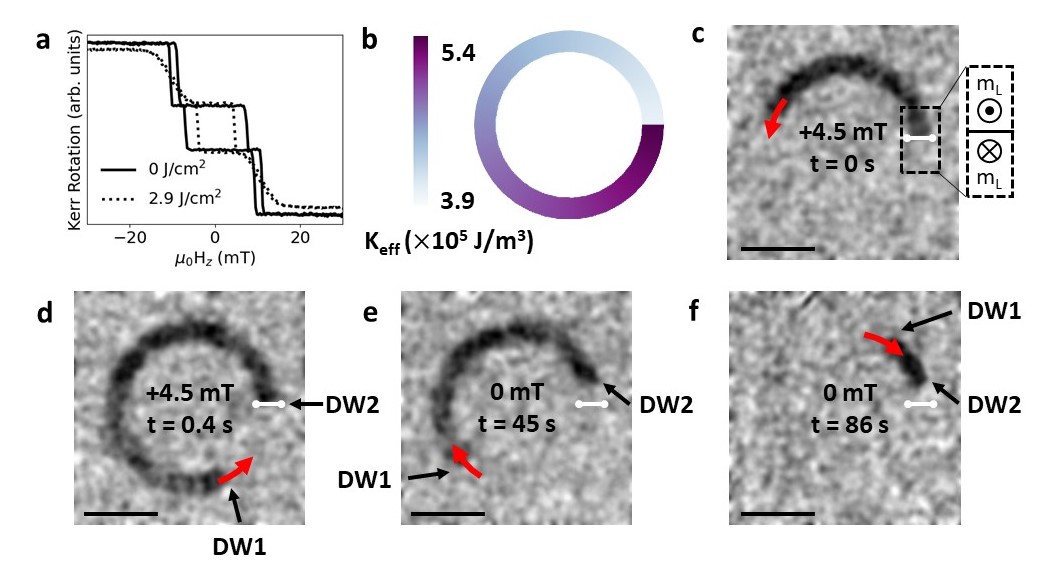} 
    \caption{\textbf{Magnetic domain wall motion along an anisotropy gradient in a SAF.} \textbf{a,} As-grown SAFs display a spin-flip transition (black solid line), where the two magnetic layers abruptly switch from antiferromagnetic to ferromagnetic alignment (at $\sim10$ mT). After laser exposure at a fluence of 2.9 \jcm, the SAF instead has a more gradual spin-flop transition (black dashed line). The standard error of the measurements shown in a are smaller than the lines used. \textbf{b,} The circular anisotropy gradient written into a continuous CoFeB-based SAF, giving a radial domain wall ratchet. \textbf{c-f,} Background-subtracted Kerr microscope images of the device (scale bar = 10 $\mu$m). The background was taken at remanence after application of a -30 mT magnetic field. The red arrows indicate the direction of domain wall motion. As the Kerr contrast of SAFs at remanence is small, the contrast was enhanced with post-processing. Unaltered images are shown in Extended Data Fig. 4. \textbf{c,} The device just at the moment a +4.5 mT magnetic field is applied. A domain nucleates at the lowest anisotropy region (labeled with a white dumbbell), with the domain wall configuration shown on the right. Here, m$_\text{L}$ refers to the magnetization of the lower magnetic layer in the SAF, which determines the Kerr contrast. \textbf{d,} The device shortly after applying a +4.5 mT magnetic field pulse. DW1 travels along the gradient, while DW2 remains pinned at the barrier between low and high anisotropy. \textbf{e,} 45 s after the applied magnetic field returns to zero, DW1 has started retreating back down the anisotropy gradient. DW2 moves counterclockwise slightly, possibly due to sample inhomogeneity which leads to a small region of reduced remanent magnetization. \textbf{f,} At 86 seconds after field pulse, DW1 has continued to move spontaneously down the anisotropy gradient towards DW2, where the domain nucleated.}
    \label{fig:safAnnealing}
\end{figure}

To monitor tuning of magnetic anisotropy in the CoFeB-based SAFs due to DWLA, we record pMOKE hysteresis loops of regions annealed with varying laser fluences. We observe a shift in the pMOKE loop from a spin-flip SAF, where there is a sharp switching event from antiferromagnetic to ferromagnetic alignment at $\mu_0H_\text{RKKY}$, to a spin-flop SAF, where switching from antiferromagnetic to ferromagnetic alignment occurs gradually, at a fluence of 2.9 \jcm (seen in Fig. \ref{fig:safAnnealing}a). If laser annealing modified the anisotropy of the top magnetic layer only, an increase of the remanent magnetization would be expected. However, the remanent magnetization is unaffected, suggesting the anisotropy of both magnetic layers are reduced simultaneously (see Extended Data Fig. 5a). A further increase in laser fluence (to $\geq3.4$ \jcm) leads to a loss of PMA.

The grayscale and complex design capabilities of DWLA provide an opportunity to make devices with much greater freedom than with existing thin film modification methods. Specifically, gradients along curved paths and around corners can be created. As a proof of concept, we fabricate a circular domain wall track where the laser fluence linearly decreases along the track such that the anisotropy gradient is always parallel to the track (seen in Fig. \ref{fig:safAnnealing}b). Such a structure is impossible to fabricate with most other methods except ion irradiation, which suffers from difficulty in maintaining constant doses radially \cite{franken_shiftregister}. The linear increase in laser fluence leads to a linear decrease in the PMA of the SAF, with $\Delta K_{\text{eff}}=1.5\times10^5$ J/m$^3$ (see Extended Data Fig. 5b). The device is initialized with a -30 mT magnetic field to a down-up state (referring to the magnetization direction in the bottom and top magnetic layer, respectively). When a positive magnetic field is applied, an up-down domain nucleates in the lowest anisotropy region (indicated by the white dumbbell in Fig. \ref{fig:safAnnealing}c), and one domain wall (DW1) propagates counterclockwise along the circular anisotropy gradient. Due to the anisotropy energy barrier, the other domain wall (DW2) is prevented from moving clockwise (see Fig. \ref{fig:safAnnealing}d). When the field is removed, DW1 retreats to the area with the lowest anisotropy, moving spontaneously down the anisotropy gradient, seen in Fig. \ref{fig:safAnnealing}e,f. Such “resetting functionality'' is promising for providing a feedback mechanism for domain wall logic, which is essential for the development of in-memory computing, or as a resetting mechanism for neuromorphic computing \cite{luo2020current, mah_domain_2021}.

\section{Discussion}
There are five key advantages that direct-write laser annealing offers over pre-existing methods for local modification of materials properties:
\begin{enumerate}
    \item Ease of use: DWLA is performed using commercial equipment designed for 2D and 2.5D photolithography. The desired exposure profile is straightforwardly created using CAD software and interpreted by the tool. 
    \item Grayscale capability: Unlike magnetic property gradients produced by composition or thickness wedges, gradients produced by DWLA can have arbitrary 2D shapes and energy profiles. While light ion irradiation offers capabilities similar to DWLA, beam instability may impact exposure precision over large areas, and the required instrumentation is less readily available. Hence, DWLA significantly expands the possibilities for creating mesoscopic, arbitrary profiles in magnetic potential energy that extend over large areas. This flexibility makes DWLA ideal to couple with computational approaches to control magnon, phonon, or photon propagation \cite{wang2021inverse, dorn2023inverse}.
    \item Modification through the whole heterostructure: Unlike selective oxidation by oxygen plasma and thermal scanning probe lithography, thermal energy is deposited throughout the heterostructure, allowing alteration of magnetic properties through the entire thickness of a film rather than just at the surface. 
    \item Speed: For 180 nm feature sizes (see Extended Data Fig. 6), a $100\times100$ $\mu$m$^2$ area can be exposed in under one minute. This is an important practical advantage over techniques like thermal scanning probe lithography, which typically requires more processing time \cite{levati2023phase}. The high speed of DWLA also means that large areas (mm to cm scale) can be exposed with high uniformity. It is further anticipated that using multiple laser sources in parallel could facilitate even faster exposure speeds. 
    \item Generality: Because DWLA can be implemented in any thin film modified by heat-induced crystallization, oxidation, alloying, or diffusion, it is relevant to a wide variety of material systems beyond magnetic materials. For example, since non-uniform charge currents at corners of heavy metal conduits can induce field-free spin-orbit torque switching \cite{kateel2023field}, a precise control of switching characteristics could be envisioned from a spatial tuning of resistivity. The possibility to create nanoscale gradients in the refractive index of a material extends the applicability of DWLA to the field of photonic metamaterials \cite{jin2019gradient}. Very recently, DWLA was used to create photonic circuits by modifying the crystal structure of a thin film in a binary manner \cite{wu2024freeform}. Our grayscale technique dramatically expands the possibilities for photonic devices, and can be extended to any photonic material whose properties can be continuously tuned by laser irradiation. While surface roughness gradients created using multiple chemical and microfluidic patterning steps have been used to control the properties of cultured cells \cite{zhou2015_roughness}, DWLA offers a path to obtain similar effects with a substantially simpler process. 

\end{enumerate}

\section{Conclusions}
We have demonstrated how intricate magnetic energy landscapes can be created on demand using direct-write laser annealing. This opens up the possibility of realizing theoretically proposed device designs that could not be fabricated using existing methods. Such fine control of the energy landscape, both in terms of pattern and energy scale, will make it possible to characterize with high precision how spin waves and magnetic domain walls interact with variations in magnetic properties. This will lead to new schemes for in-memory computation, data storage, and sensing. Beyond the field of magnetism, direct-write laser annealing provides a path to locally control and exploit the properties of any thin film that transforms in response to heat, ushering in further applications across many fields.

\section{Methods}
\subsection{Synthesis}
All films were deposited on Si substrates with a 300 nm-thick thermal oxide coating using magnetron sputtering in a 3 mTorr Ar partial pressure. The ATC Orion sputtering tool (AJA International, Inc.) used had a base pressure of $4\times10^{-8}$ Torr. MgO was deposited with rf sputtering; all other layers were deposited with dc sputtering. The full composition of the films are detailed below, where all thicknesses are in nm: 
\begin{itemize}
    \item CoFeB films, with the structure Ta(5)/ Co$_{40}$Fe$_{40}$B$_{20}$(1.3)/ MgO(1.5)/ Ta(5) were sputtered at powers of 80 W, 40 W, and 150 W, respectively. 
    \item CoGd films of the structure Ta(2)/ Pt (5)/ Co$_{70\%}$Gd$_{30\%}$ (5.5)/ Ta (3) were deposited using a co-sputtering technique, in which the Co (Gd) was deposited using a sputtering power of 50 W (24 W).
    \item The Co/Cr/Co SAFs have the composition Ta(2)/ Pt(3)/ Co(0.7)/ Cr(1)/ [Co(0.7)/ Pt(1)]$_{3}$/ Pt(5). All layers were deposited using a sputtering power of 100 W. 
    \item The CoFeB/Pt-based SAFs have the structure Ta(5)/ [Pt(1.5)/ CoFeB(0.4)]$_3$/ Ru(1.1)/ [CoFeB(0.4)/ Pt(1.5)]$_2$. Ta and Ru were deposited at 80 W, CoFeB was deposited at 40 W, and Pt was deposited at 100 W. 
\end{itemize}

After treatment with DWLA, the Co/Cr/Co sample was patterned to a 2 $\mu$m-wide spiral feature (as shown in Fig. \ref{fig:SAFJB_2}a) using conventional UV photolithography and ion beam etching techniques.

\subsection{Direct-write laser annealing}
A Heidelberg DWL66+ direct-write photolithography system with 405 nm CW illumination was used to perform DWLA. In the “Advanced Grayscale” configuration, this platform provides 256 levels of grayscale exposure intensity, which allows for quasi-continuous variations in magnetic properties. Design files made using standard CAD programs modulate the power of the laser as it is raster-scanned over the film. The “Hi-Res” write head used throughout this work offers a 50 nm address grid with feature sizes of 180 nm (see Extended Data Fig. 6). The maximum laser fluence possible is $\sim 31$ k\jcm. Based on the published write speed of 3 mm$^2$/minute, we estimate that each pixel in the address grid is exposed to the laser for 50 ns. 

The substrate on which a film is grown can dramatically impact the laser fluence required to induce transformations in the physical properties. As the laser annealing process is thermal in nature, this variance can be explained by the differing thermal conductivities or optical absorptivities of the substrates. Silicon with a 300 nm-thick thermal oxide coating is used for all experiments in the paper, as the low thermal conductivity of SiO$_2$ (1.3 W/m$\cdot$K) reduces the laser fluence needed to obtain significant temperature increases. Similar modifications in magnetic properties can be achieved for films deposited on Si$_3$N$_4$ (55 W/m$\cdot$K)\cite{dow2017thermal} and Si (140 W/m$\cdot$K)\cite{li2017thermal} substrates, but much higher fluences must be used. We also found that different SiO$_2$ wafers required slightly different fluences to achieve the same effects, indicating a strong sensitivity to the thermal properties of the substrate. As the maximum fluence ($\sim 31$ k\jcm) was not required to modify films grown on any of the substrates mentioned above, DWLA is suitable for materials with higher thermal conductivities or lower optical absorptivities.

\subsection{Magneto optic Kerr effect measurements}
Magneto optic Kerr effect (MOKE) hysteresis loops were measured using a NanoMOKE magnetometry system manufactured by Durham Magneto Optics, Ltd., which employs a 660 nm laser. The pMOKE hysteresis loops were collected from areas that were direct-write laser annealed using the indicated laser fluence. Imaging of magnetic domain structures was performed using a MOKE microscopy platform with wide-spectrum LED illumination manufactured by Evico Magnetics GmbH. As is common in MOKE microscopy, the contrast of MOKE images was enhanced using a background subtraction procedure in which an image collected during exposure to a saturating magnetic field was subtracted from the recorded image.  

\subsection{Magnetic characterization}
A Quantum Design magnetic property measurement system (MPMS) equipped with superconducting quantum interference device - vibrating sample magnetometry (SQUID - VSM) was used to measure the saturation magnetization of selected continuous films subjected to uniform laser exposure. 

\section{End Notes}
\subsection{Acknowledgements}
JAB acknowledges funding from the European Union’s Horizon 2020 research and innovation programme under the Marie Skłodowska-Curie grant agreement No 884104 (PSI-FELLOW-III-3i). LJR acknowledges support from the ETH Zurich Postdoctoral Fellowship Program 22-2 FEL-006. We thank J. Moritz Bosse for assistance with furnace annealing experiments. We thank the staff of the cleanroom facilities at the Laboratory for Nano and Quantum Technologies at the Paul Scherrer Institute for technical support. 

\subsection{Author contributions}
JAB and LJR synthesized samples, carried out experiments, and performed analysis. KM synthesized SAF samples and carried out characterization experiments. JAB, LJR, AH, and LJH prepared the manuscript.

\bibliography{refs}

\clearpage
\section{Extended Data}
\begin{figure}[htb!]
    \centering
    \includegraphics[width=\textwidth]{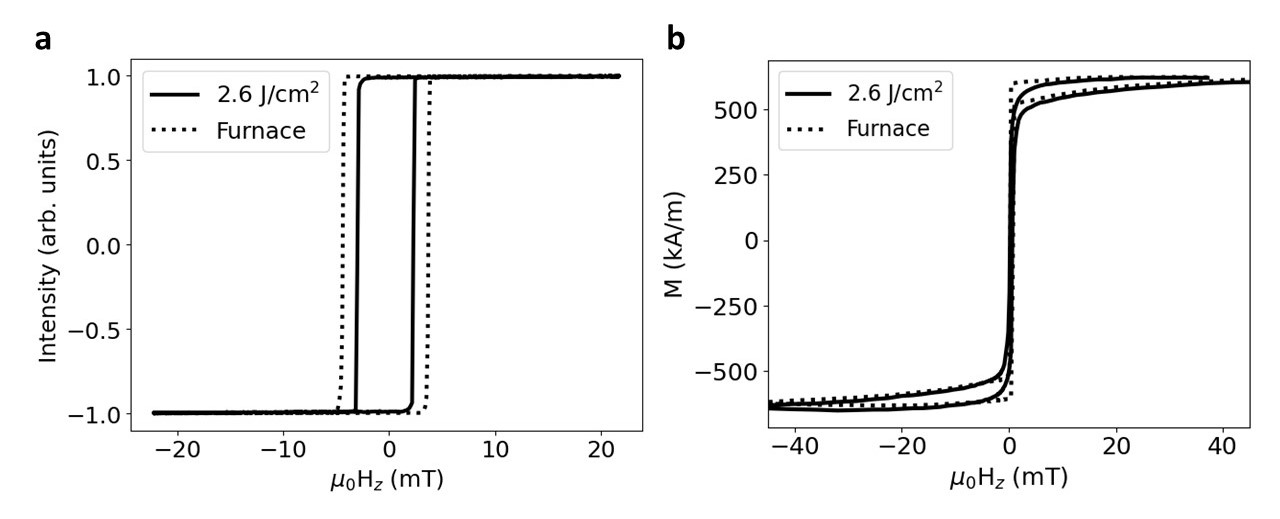}
    \caption*{\textbf{Extended Data Fig.1 Comparison of CoFeB magnetic properties for laser annealing and furnace annealing.} A laser annealed CoFeB film is compared to an identical furnace annealed sample (annealed in a tube furnace at atmosphere for 1 hour at 300 $^\circ$C with a temperature ramp rate of 20 $^\circ$C/minute). \textbf{a,} Polar MOKE measurements of the two samples reveals similarly square hysteresis loops, with only a small difference in the coercive field. \textbf{b,} Out-of-plane SQUID - VSM measurements for both a laser annealed and a furnace annealed sample. The saturation magnetization is the same for both samples, indicating that the two techniques give an equivalent transformation of magnetic properties in CoFeB. The standard error of the measurements shown are smaller than the lines used.}
    \label{fig:CFBfurnace}
\end{figure}
\clearpage

\begin{figure}[htb!]
    \centering
    \includegraphics[width=1.0\linewidth]{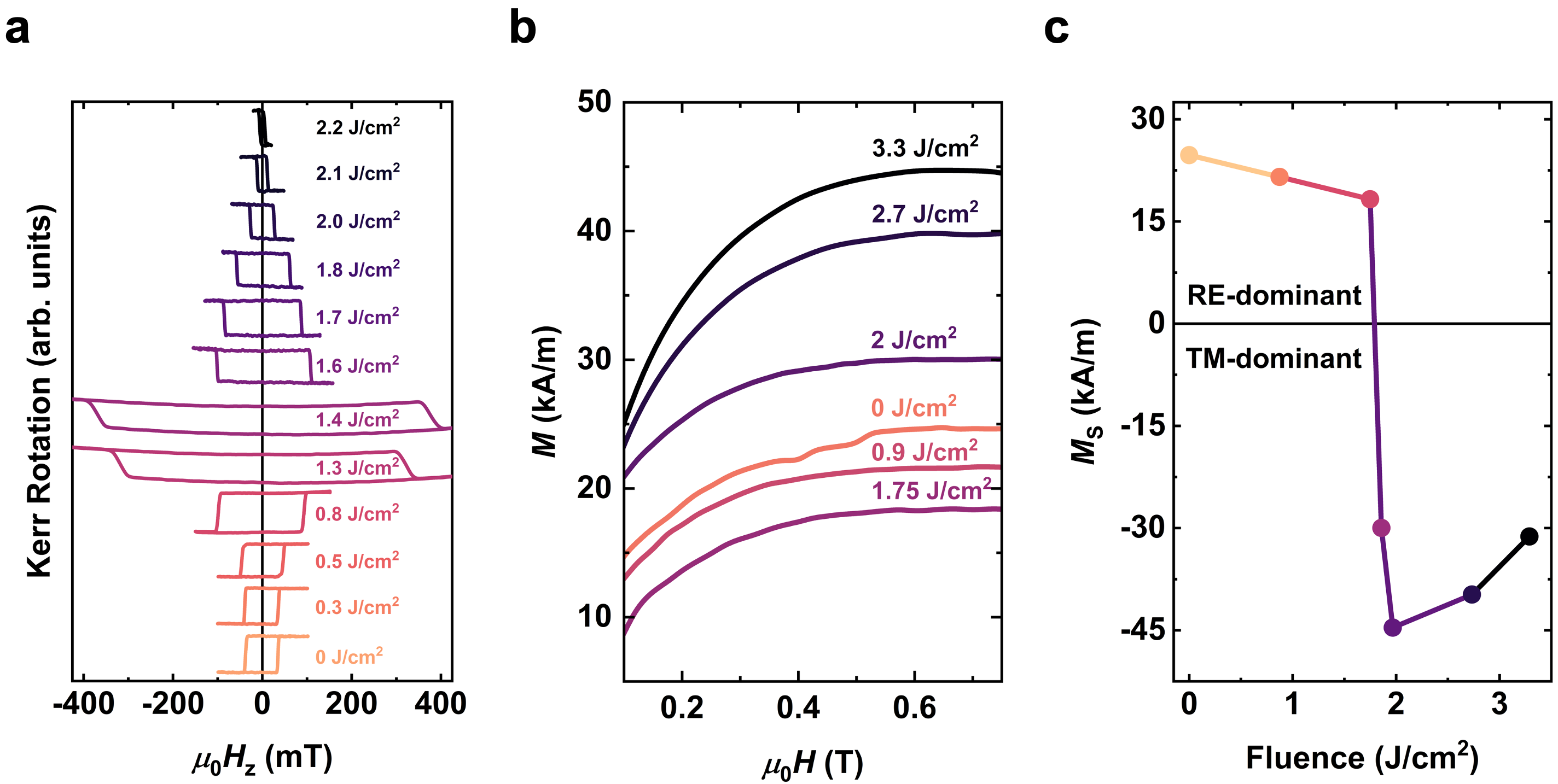}
    \caption*{\textbf{Extended Data Fig.2 Magnetic characterization of CoGd films.} \textbf{a,} Room temperature pMOKE hysteresis loops for $100\times100$ $\mu$m$^2$ regions of a CoGd film exposed using various laser fluences. \textbf{b,} Room temperature, in-plane SQUID - VSM data for continuous CoGd films that were uniformly exposed using the indicated laser fluences. \textbf{c,} The saturation magnetization as a function of laser fluence extracted from \textbf{b}. The standard error of the measurements shown are smaller than the symbols or lines used.} 
    \label{fig:enter-label5}
\end{figure}
\clearpage

\begin{figure}[htb!]
    \centering
    \includegraphics[width=1\linewidth]{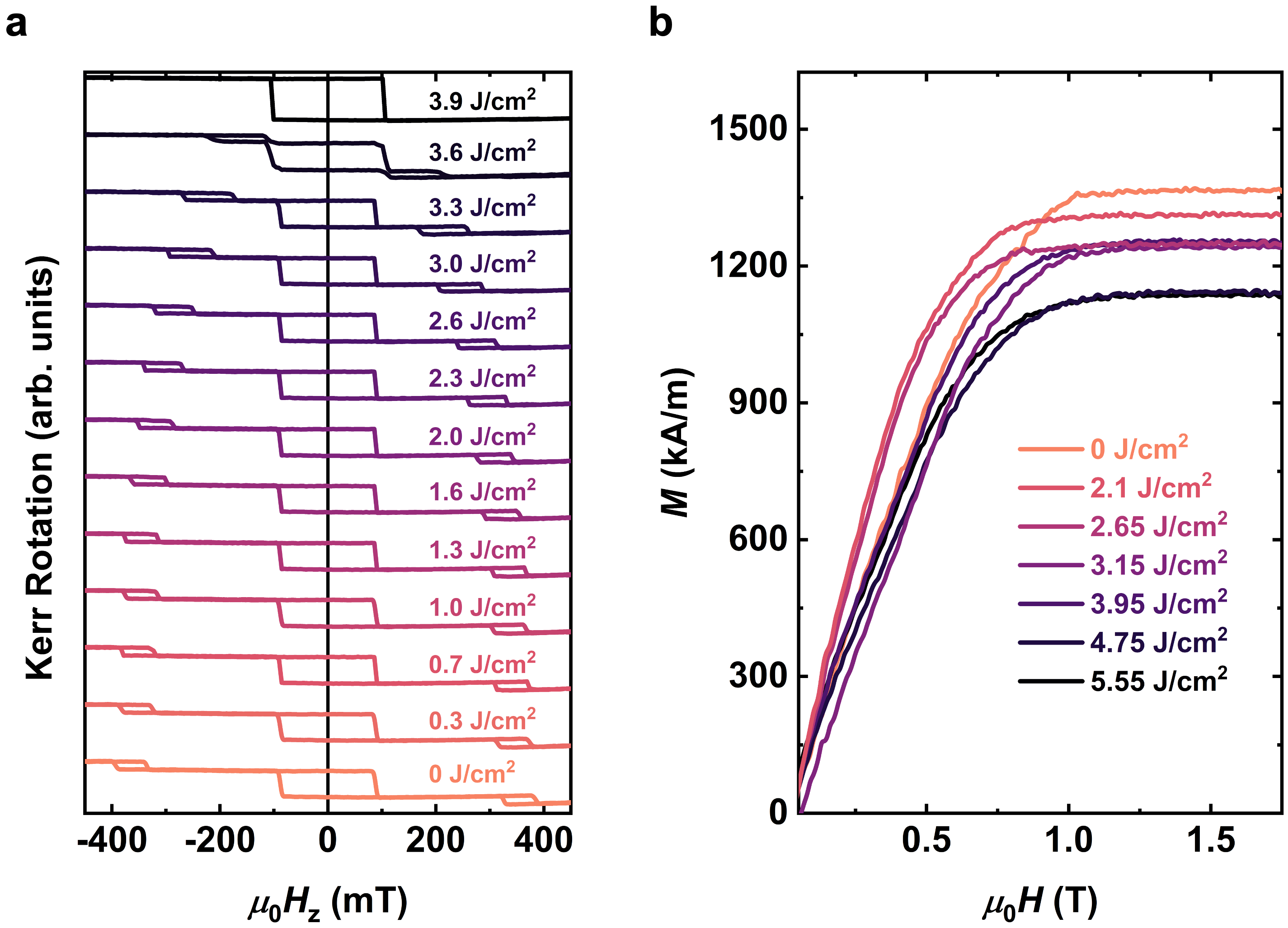}
    \caption*{\textbf{Extended Data Fig.3 Magnetic characterization of Co/Cr/Co films.} \textbf{a,} Room temperature, pMOKE hysteresis loops for $100\times100$ $\mu$m$^2$ regions of a Co/Cr/Co film exposed using the indicated laser fluences. \textbf{b,} Room temperature, in-plane SQUID - VSM measurements of continuous Co/Cr/Co films that were uniformly exposed using the indicated laser fluences. The standard error of the measurements shown are smaller than the lines used.}
    \label{fig:enter-label3}
\end{figure}
\clearpage

\begin{figure}[htb!]
    \centering
    \includegraphics[width=\linewidth]{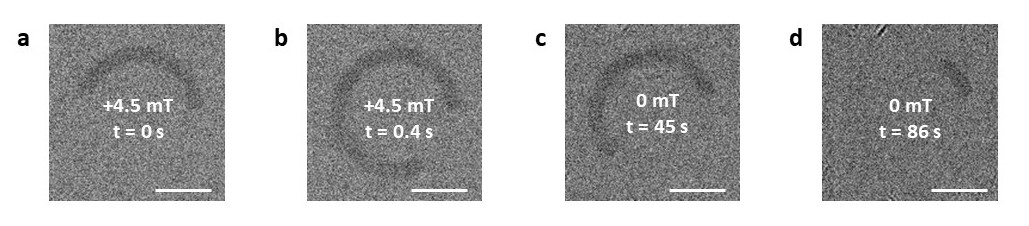}
    \caption*{\textbf{Extended Data Fig.4 Kerr microscope images of domain wall motion around a loop.} The original unprocessed Kerr microscope images seen in Fig. 4c-f (scale bar = 10 $\mu$m). For images in the main text, a Gaussian blur was applied and contrast was enhanced.}
    \label{fig:lr-saf-dw}
\end{figure}
\clearpage

\begin{figure}[htb!]
    \centering
    \includegraphics[width=\linewidth]{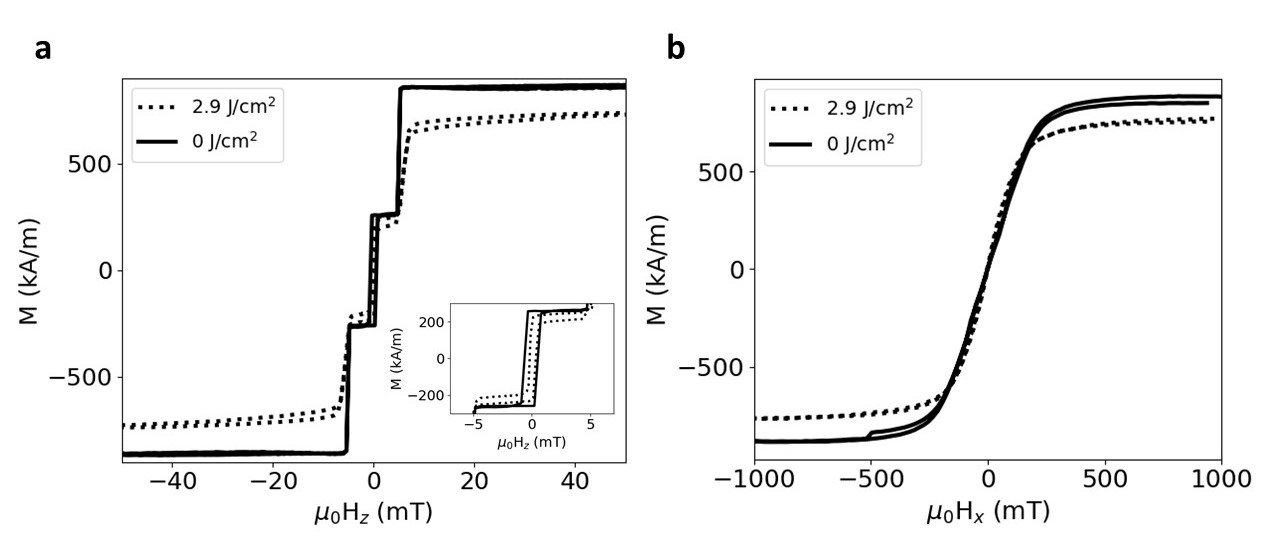}
    \caption*{\textbf{Extended Data Fig.5 Magnetometry of CoFeB/Pt-SAF films.} Room temperature SQUID-VSM measurements of two identical SAF films, one of which has been laser annealed at a fluence of 2.9 J/cm$^2$. \textbf{a,} Out-of-plane magnetometry reveals that the saturation magnetization of the laser annealed sample is $\sim10$ \% lower than that of the unannealed film. Inset: the remanent magnetization of the laser annealed film is also slightly lower than that of the unannealed film. If only the anisotropy of the top layer was reduced by DWLA, an increase of the remanent magnetization would be expected, as a canted moment in the top magnetic layer would not compensate as much of the moment from the bottom magnetic layer. The nearly identical remanent magnetization indicates that the thermal energy is being evenly deposited in both the top and bottom magnetic layers. \textbf{b,} In-plane magnetometry of the laser annealed and unannealed SAF films. The effective anisotropy field and saturation magnetization is slightly lower for the annealed sample. The standard error of the measurements shown are smaller than the lines used.}
    \label{fig:lr-saf-squid}
\end{figure}
\clearpage

\begin{figure}[htb!]
    \centering
    \includegraphics[width=0.4\linewidth]{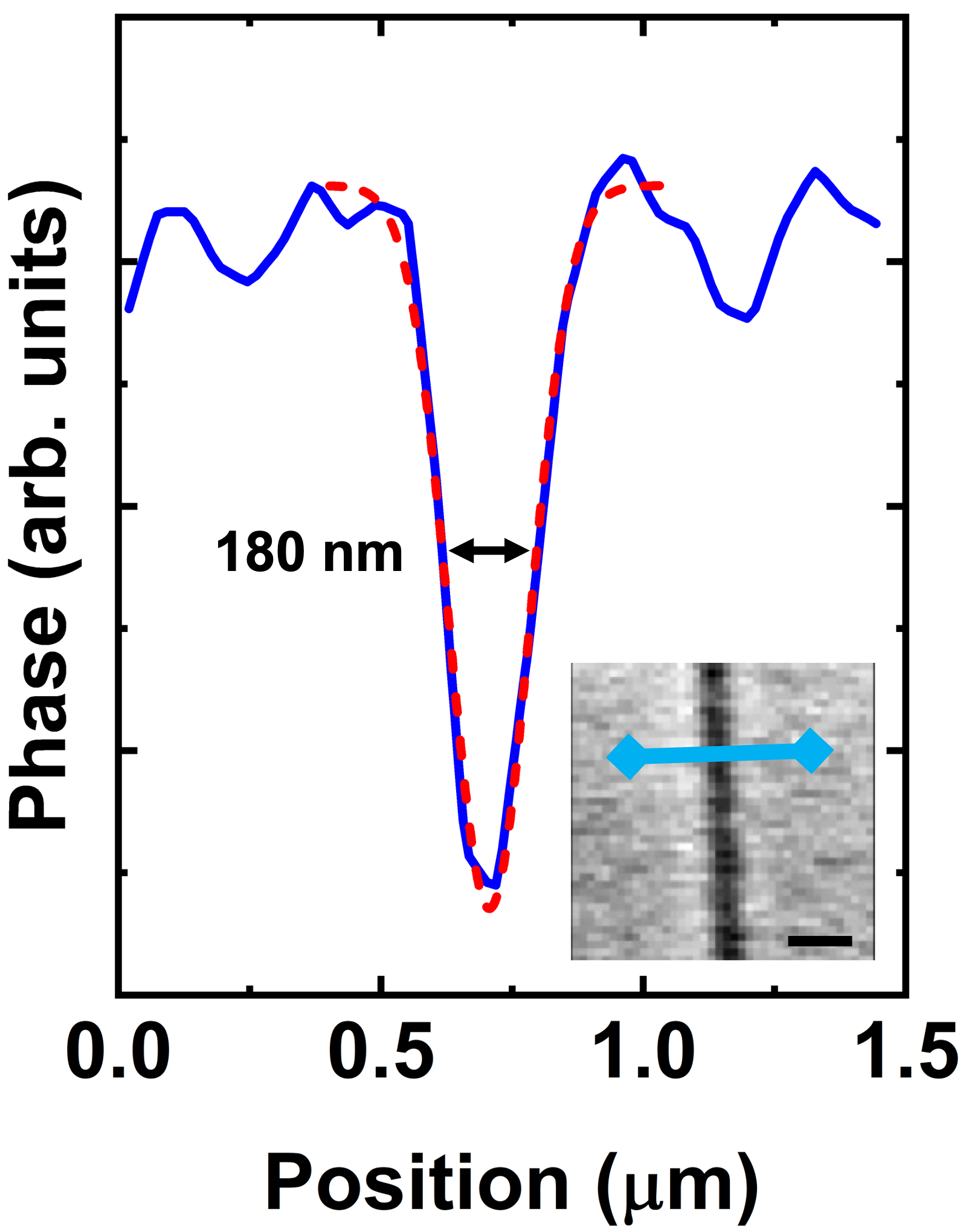}
    \caption*{\textbf{Extended Data Fig.6 Determination of DWLA feature size using magnetic force microscopy.} Line profile (solid blue line) and Gaussian fit (dashed red line) of magnetic force microscopy (MFM) contrast for a Co/Cr/Co SAF film in which DWLA (with a laser fluence of 6.5 \jcm) was used to create a ferromagnetically coupled region (dark region in the inset image), which exhibits a full-width half-minimum of approximately 180 nm. The profile was extracted along the blue line shown in the inset MFM image (scale bar = 500 nm). The nominal width of the exposed area was 50 nm in the design.} 
    \label{fig:mfm-linecut}
\end{figure}
\clearpage
\end{document}